\documentclass[aps,twocolumn,nofootinbib,preprintnumbers,superscriptaddress]{revtex4-1}

\usepackage{amsmath, mathrsfs, amssymb,amsfonts,amsthm,graphicx, epsf, dcolumn, yfonts}
\usepackage{latexsym}
\usepackage[colorlinks=true, pdfstartview=FitV, linkcolor=black, citecolor=black, urlcolor=black]{hyperref}

\usepackage{color}
\usepackage{slashed}
\newcommand{\be}{\begin{equation}}
\newcommand{\ee}{\end{equation}}
\newcommand{\bea}{\begin{eqnarray}}
\newcommand{\eea}{\end{eqnarray}}
\newcommand{\bwt}{\begin{widetext}}
\newcommand{\ewt}{\end{widetext}}

\def\ie{{\it i.e.~}}

\def\k{\kappa}

\begin{document}

\title{Aspects of Current Correlators in Holographic Theories with Hyperscaling Violation}
\author{Mohammad Edalati and Juan F. Pedraza}
\affiliation{Theory Group, Department of Physics and Texas Cosmology Center, University of Texas at Austin, Austin, TX 78712, USA}

\preprint{UTTG-17-13}
\preprint{TCC-014-13}

\begin{abstract}

We study the low energy and low momentum behavior of current correlators in a class of holographic zero-temperature finite density critical theories which do not respect the hyperscaling relation. The dual holographic description is assumed to be given by probe D-branes embedded in background geometries characterized by a dynamical critical exponent $z$ and a hyperscaling violation exponent $\theta$. We show that a subset of these theories with $1\leq z<2(1-\theta/d)$ exhibit a stable linearly-dispersing mode in their low energy spectrum of excitations. This mode, which appears as a pole in the retarded correlators of charge density and longitudinal currents, has some characteristics similar to that of the zero sound in Fermi liquids. Given some reasonable assumptions, we argue that the class of theories with $\theta =d-1$ that logarithmically violate the area law in the entanglement entropy in a manner reminiscent of theories with Fermi surfaces, does \emph {not} exhibit a zero sound-like mode in the low energy spectrum of the probe sector. Furthermore, utilizing the holographic Wilsonian approach, we explicitly show that such a mode has a natural interpretation as a Goldston boson arising from the spontaneous breaking of a specific symmetry.

\end{abstract}

\maketitle

\section{Introduction}

In close proximity of a critical point in a second-order phase transition, observables of the theory typically exhibit scaling behavior given by a plethora of critical exponents. These exponents are not all independent; they satisfy a set of relations amongst themselves \cite{Goldenfeld:1992qy}. One of these relations, known as the Josephson scaling law, is distinguished from the other scaling relations in that the dimensionality of space, $d$,  appears explicitly. As such, it is an example of the so-called hyperscaling relations \cite{Widom1965} whose proofs require further assumptions\footnote{Such as, sufficiently close to the critical point the free energy per unit volume scales naturally with the correlation length,  provided that the correlation length is the only relevant length scale.}  in addition to the usual scaling arguments. It is rather well known that the Josephson scaling law, hereafter referred to as simply the hyperscaling relation, could fail in the presence of dangerously irrelevant couplings \cite{WegnerRiedel1973, M.E.Fisher1974}. Typical examples of critical theories which do not respect the hyperscaling relation include those in their upper critical dimension, or some even below their critical dimension, as in the random-field Ising model \cite{D.S.Fisher1986}.

In the context of holography, strongly coupled critical theories with a violation of the hyperscaling relation could be modeled  using rather simple gravitational systems. The gravity solution in such cases is characterized by two parameters \cite{Ogawa:2011bz,Gouteraux:2011ce}, $z$ and $\theta$, interpreted in the dual critical theory as representing the dynamical critical exponent and the hyperscaling violation exponent \cite{Huijse:2011ef}, respectively. An intriguing feature of these holographic systems is that for $\theta = d-1$ the entanglement entropy in the dual theory exhibits a logarithmic violation of the area law \cite{Ogawa:2011bz, Huijse:2011ef} with the structure of the logarithmic term being similar to those one expects to observe in theories with Fermi surfaces \cite{Wolf:2006zzb, Gioev:2006zz, Swingle:2009bf, Swingle:2010yi, Zhang:2011}. Moreover, if considered at finite temperature, theories with $\theta=d-1$ exhibit an entropy whose dependence on temperature is as if the excitations are effectively one dimensional \cite{Gouteraux:2011ce,Huijse:2011ef}. This is another feature which suggests that these holographic theories may have some sort of `fermionic' character, even though there are no explicit fermionic fields in their dual bulk  description.

In this paper we study the low energy behavior of some observables in these theories at zero temperature, but finite charge density.  Observables of interest to us are mainly the correlation functions of currents in the longitudinal channel. One of our motivations for computing these particular observables is to further investigate their fermionic imprints (or lack thereof) for the case $\theta=d-1$. One such investigation is the fate of a zero sound-like mode in these theories which, if exists, will show up as a pole in the retarded correlators of charge density and longitudinal currents. Such a mode, with properties similar to the zero sound mode in Fermi liquids, have been shown to exists in a number of other zero-temperature finite density holographic theories \cite{Karch:2008fa, Kulaxizi:2008kv, Kulaxizi:2008jx, Edalati:2010pn, HoyosBadajoz:2010kd, Lee:2010ez, Davison:2011ek}. Another motivation, which is actually not specific to the type of observables we are considering here, is that  the study of correlation functions in these theories and the classification of their short and long distance behavior as a function of $z$ and $\theta$ is an interesting topic on its own right. Such studies will hopefully help us sharpen the similarities as well as the differences between holographic theories with hyperscaling violation and the those that arise in the context of critical phenomena  in condensed matter physics.

Our analysis of the correlators shows that at low frequency and low momentum (compared to the chemical potential) there exists a stable zero sound-like mode in the low energy spectrum of the probe sector for those theories where the dynamical critical exponent and the hyperscaling violation exponent satisfy the condition $1\leq z<2(1-\theta/d)$. The zero sound-like mode that we find has a dispersion relation of the form $\omega = v_0 k -i \Gamma k^{2(1-\theta/d)/z}$ where $v_0$ and $\Gamma>0$ are some constants whose expressions are given in the bulk of the paper. Given some reasonable assumptions described in the next section, we argue that the case $\theta=d-1$ is not included in the above range. Our finding in this particular case seems to be consistent with the claim in \cite{Hartnoll:2012wm} that the low frequency spectral density of transverse currents is negligible at finite momentum.  In fact, as explained in \cite{Hartnoll:2012wm}, the lack of appreciable low energy, finite momentum excitations is not specific to the case $\theta=d-1$ but applies, rather generally, to all allowed values of $z$ and $\theta$ (as long as the $z\to \infty$ limit is excluded). This behavior then poses a challenge as to what the true nature of the zero sound-like mode is in cases where we find it in the low energy spectrum. In fact, following \cite{Nickel:2010pr}, we show that such a holographic mode, instead of being attributed to a Fermi surface, could be naturally taught of as being a Goldston mode coming from the spontaneous breaking of a U(1)$\times$U(1) (one being a global symmetry associated with a conserved charge and the other being a gauge `symmetry' due to an emergent Abelian gauge field) down  to the diagonal U(1).

The organization of the paper is as follows.  In section \ref{sec:2} we review salient features of the holographic setup describing theories with hyperscaling violation. We then consider these theories at finite density for a U(1) charge by embedding a D-brane in the bulk geometry.
In section \ref{sec:3} we analytically compute the retarded correlators of the charge density and longitudinal current operators at low frequency and low momentum, where  we establish the existence of a stable linearly-dispersing mode in the low energy spectrum for a subset of the ($z$,$\,\theta$) parameter space, namely for when $1\leq z<2(1-\theta/d)$. In section \ref{sec:4} we show that this mode could be given an effective description and be interpreted as a Goldston mode of a spontaneous symmetry breaking. Finally, we conclude in section \ref{sec:5} with some discussions as well as directions for future work.\

{\it Note added: } While we were at the final stages of writing up our paper we became aware of two papers \cite{Pang:2013ypa, Dey:2013vja} whose results overlap with ours. The results in \cite{Pang:2013ypa} has substantial overlap with ours while \cite{Dey:2013vja} considers only a special limit of the setup we consider here, where one takes $z \to \infty$ with $\eta =-\theta/z>0$ held fixed. As we show later, our results agree with those of \cite{Pang:2013ypa} in this limit.

\section{Holographic Theories with Hyperscaling Violation\label{sec:2}}

\subsection{Bulk Solution}

Consider a $(d+2)$-dimensional bulk theory with a negative cosmological constant whose metric is of the form
\begin{align}\label{bckgr2}
ds^2 &\equiv \,g_{\mu\nu}dx^\mu dx^\nu\nonumber\\
&=\,\frac{1}{r^{2(1-\theta/d)}}\left(-\frac{dt^2}{r^{2z-2}}+dr^2+d\vec{x}^2\right),
\end{align}
where $z$ and $\theta$ are, for now, some arbitrary constants and $d$ is the number of spatial dimensions, $d\vec x^2 = dx_1^2 +\cdots + dx_d^2$. For $\theta=0$, and $z\neq 1$, the metric \eqref{bckgr2} is that of Lifshitz \cite{Kachru:2008yh,Singh:2010zs,Singh:2012un}, and is invariant under the scale transformation
\begin{align}\label{scalings}
t \to \lambda^z t,\,\,\,\, \vec{x} \to \lambda \vec{x},\,\,\,\, r \to \lambda r\,.
\end{align}
For any non-zero $\theta$, on the other hand, the above metric, which is conformally related to a Lifshitz metric, transforms covariantly
\begin{align}
ds \to \lambda^{\theta/d} ds\,,
\end{align}
under the scale transformation \eqref{scalings}. Needless to say that for $\theta=0$ and $z=1$ the metric is that of AdS in the Poincar\'{e} patch. Later, we will discuss an interesting limit of the metric which makes it conformal to AdS$_2\times \mathbf{R}^d$.

By now, there are quite a few number of bulk theories in the literature which give rise to a metric of the form \eqref{bckgr2}.  From the perspective of a bottom-up approach, such a metric can arise as a solution of some effective Einstein-Maxwell-dilaton theories \cite{Ogawa:2011bz, Huijse:2011ef} with a schematic Lagrangian of the form
\begin{align}
{\cal L} = {\cal R} - e^{\gamma(z,\theta) \phi}F^2 - (\partial \Phi)^2 +V_0^2 e^{\delta(z,\theta)\phi}.
\end{align}
Here, $\phi$ is a `diatonic' scalar whose non-trivial profile in the bulk would break the scale invariance\footnote{See also \cite{Gubser:2009qt,Goldstein:2009cv,  Iizuka:2011hg, Charmousis:2010zz} for earlier studies of similar Einstein-Maxwell-dilaton theories.}. In a top-down approach, the metric \eqref{bckgr2} also arises as a solution describing the low-energy behavior of a number of string/M-theory constructions in certain limits \cite{Dong:2012se, Narayan:2012hk, Dey:2012tg, Dey:2012rs, Perlmutter:2012he, Dey:2012fi, Bueno:2012sd,Gath:2012pg}. For instance, it was shown in \cite{Dong:2012se} that compactifying the ten-dimensional supergravity solution of $N$ D2-branes over $\mathbf{S}^6$, and taking its near-horizon limit, results in a metric with $z=1$, $\theta =-1/3$, $d=2$ over a wide range of the radial coordinate $r$.

While it is very interesting to explore the landscape of gravitational theories giving rise to geometries of the form \eqref{bckgr2}, in this paper,  we would like to keep the discussion general and will not restrict ourselves to a particular model in the bulk. We do, however, assume that the solution of the bulk theory satisfies some rather general constraints, such as the null energy condition, and is thermodynamically stable if placed at a finite temperature. Indeed, imposing the null energy condition, $T_{\mu\nu} n^\mu n^\nu \geq 0$ (with $n^\mu$ being any null vector) restricts the set of allowed values of $z$ and $\theta$ to the ones that satisfy \cite{Ogawa:2011bz, Dong:2012se}
\bea\label{NEC}
&&(d-\theta)\left[d(z-1)-\theta\right]\geq 0\,,\nonumber\\
&&(z-1)(d+z-\theta)\geq 0\,.
\eea
As is evident from the first inequality above, for the case $\theta = 0$, one deduces that $z \ge 1$; a result which was previously obtained in the studies of Lifshitz spacetimes \cite{Kachru:2008yh,Koroteev}. For any non-vanishing $\theta$, on the other hand, the above inequalities could be satisfied with $z<1$. Indeed, either $0<z<1$ together with $\theta \ge d+z$, or $z<0$ together with $\theta>d$ would satisfy both of them. The former case leads to a thermodynamic instability once the zero-temperature metric \eqref{bckgr2} is generalized to a finite-temperature black hole metric \cite{Dong:2012se}. We will also exclude the latter case for reasons discussed in \cite{Dong:2012se}.
Thus, in addition to obeying these two inequalities, hereafter we also assume that $\theta \leq d$. This further restriction combined with the two inequalities in \eqref{NEC} then simply implies  that $z\geq 1$. In summary, we will only consider the class of theories for which $\theta \leq d$ and $z\geq 1$. Note that theories with $z=1$ belong to this class.

\subsection{Boundary Theory}

In the framework of the AdS/CFT correspondence, a ($d+2$)-dimensional bulk theory with a metric of the form \eqref{bckgr2}\footnote{There will also be a compact space $X$ if the bulk theory is viewed as coming from a compactification of a UV complete theory,  such as string theory. The compact space will be mentioned briefly around equation (\ref{embedding}) but will otherwise play no role in our discussion.} describes in the boundary a ($d+1$)-dimensional strongly-coupled zero-temperature quantum field theory with a dynamical critical exponent $z$. Moreover, as argued in \cite{Huijse:2011ef}, for $\theta\neq 0$ the dual theory violates the hyperscaling relation with the amount of violation being characterized by $\theta$, the so-called hyperscaling violation exponent \cite{D.S.Fisher1986}.

In our coordinate system, the $r \to \infty$ and $r\to 0$ regions in the bulk map to the IR and the UV of the dual theory (recall that we are assuming $\theta \leq d$), respectively. Having said that, it is important to recognize the limitations of such solutions in terms of describing the dual field theories very deep in the IR, or far in the UV. In fact, in the bulk theories  where a metric of the form \eqref{bckgr2} is realized, there is also a dilatonic scalar field (in addition to other bulk fields)  whose profile typically blows up as $r \to 0$ or $r \to \infty$ is approached. This behavior presumably signals a significant modification of the bulk geometry in either of the two regions. Nevertheless, one could still think of \eqref{bckgr2} as being a reliable solution over a wide range of $r$;  $r_\Lambda \lesssim r \lesssim r_{\rm IR}$ with $r_\Lambda$ and $r_{\rm IR}$ denoting, respectively, some cutoffs beyond which the solution  \eqref{bckgr2} cannot be trusted. For  example, the dual field theory may flow from a relativistic conformal fixed point far in the UV (which is holographically represented by an AdS spacetime) to a theory that exhibits an emergent scaling with a dynamical critical exponent $z$ and a hyperscaling violation exponent $\theta$ at some intermediate scale, with a bulk description given by \eqref{bckgr2}, to another fixed point deep in the IR. In the context of Einstein-Maxwell-dilaton theories, explicit examples of bulk solutions representing such RG flows have been constructed in \cite {Ogawa:2011bz, Bhattacharya:2012zu, Kundu:2012jn}.

Another issue which may render the holographic description of such solutions problematic (especially, at arbitrarily low energies) is the existence of a generic null curvature singularity \cite{Shaghoulian:2011aa, Copsey:2012gw} in the $r\to \infty$ region of the geometry \eqref{bckgr2}. The existence of this singularity is not specific to the case $\theta\neq 0$, and also occurs in Lifshitz-like spacetimes \cite{Copsey:2010ya, Horowitz:2011gh} for which $\theta =0$. See also \cite{Harrison:2012vy, Bao:2012yt} for some discussions on resolving  such singularities. For now, we leave aside these subtleties and revisit them in later sections when we discuss the regime of validity of our results.

\subsection{Adding Charge to the Boundary Theory}

We assume that the zero-temperature boundary field theory in question, whose dual gravitational description has a metric of the form \eqref{bckgr2},  is neutral. We are, however, interested in studying the system at finite charge density. Hence, we couple this neutral sector to a finite density of charge carriers by embedding a `flavor' D-brane in the background geometry\footnote{As we alluded to earlier, the metric \eqref{bckgr2} could also be realized in effective Einstein-Maxwell-dilaton theories. In such cases, the boundary field theory is  in a state with a finite charge density. The leading asymptotic behavior of the bulk U(1) gauge field is interpreted in the boundary theory as the chemical potential for the charge density.}. Our approach here is analogous to the one taken in \cite{Hartnoll:2009ns} for adding charge degrees of freedom to quantum critical theories dual to Lifshitz backgrounds\footnote{See also \cite{Karch:2007pd} and its citations for earlier studies of probe charge degrees of freedom in a variety of holographic backgrounds.}.

To model a finite density of charge carriers interacting with this neutral sector we start by embedding a D-brane in the background geometry \eqref{bckgr2}. The dual theory now enjoys a global U(1) symmetry. One can then introduce a finite density of charge by turning on a chemical potential for it.
From the perspective of the bulk theory, this amounts to turning on a non-trivial background value for the time-component of a U(1) gauge field on the worldvolume of the  D-brane. In what follows, we treat the brane in the probe approximation. That is, we assume that its backreaction on the background geometry could be safely ignored\footnote{As discussed in \cite{Hartnoll:2009ns}, such an assumption may fail in the deep interior region of a scaling metric similar to that of \eqref{bckgr2}. This is mainly due to either the energy density of the brane becoming comparable to that of the background somewhere in the deep interior region, or the pair creation of strings in the presence of a strong worldvolume electric field making stringy effects non-negligible, or even both. We will ignore these issues in our discussions in this section, as the probe approximation will, nevertheless, be valid for a wide range of the radial coordinate.}.

The dynamics of the aforementioned D-brane (with the number of worldvolume directions equal to $p+1$) is given by the Abelian DBI action
\begin{align}\label{FlavorDBIAction}
\mathcal{S} = -T_{q} \!\!\int\!d\tau{d}^{p}\sigma\,\sqrt{-\mathrm{det}\left[{\cal G}_{ab} + {\cal F}_{ab}\right]} \,.
\end{align}
There are also Chern-Simons like terms in the D-brane action which encodes the couplings of the worldvolume fields to the background RR form fields.  Such terms will have no bearing in our subsequent discussions, and will be ignored. In the action \eqref{FlavorDBIAction}, $T_q$ is the tension of the D$p$-brane,  $\tau$ and $\sigma^a$ ($a=1,\cdots, p$) are the worldvolume coordinates,  ${\cal G}_{ab} = g_{\mu\nu}\partial_a x^\mu \partial_b x^\nu$ is the induced metric with $g_{\mu\nu}$ representing the background metric in the Einstein frame and $F_{ab} = (2\pi \alpha' )^{-1}{\cal F}_{ab}$ is the field strength of a U(1) worldvolume gauge field. Also, the factor $(2\pi\alpha')^{-1}$ is the string tension. Hereafter, we set $2\pi \alpha' = 1$.

We seek an embedding of the form \cite{Hartnoll:2009ns}
\begin{align}\label{embedding}
&\,\tau = t, \qquad \sigma^1  =x^1,\quad \cdots \quad \sigma^q = x^q, \qquad \sigma^{q+1} = r, \nonumber\\
&\{\sigma^{q+2},\cdots, \sigma^p\}=\Sigma,
\end{align}
where $\Sigma\subseteq X$ is an internal space around which the brane is wrapped. Note that in the above embedding, the integer $q\leq d$ has been introduced to allow us consider also the cases where the flavor D$p$-brane intersects the stack of `color' D-branes.
Of course, when $q<d$ it is understood that the flavor fields in the boundary field theory propagate only along some $q$-dimensional defect.
Also, we further assume that the embedding of the worldvolume coordinates of the brane in the directions of $\Sigma$ (namely, $\sigma^{q+2}$, $\cdots$, $\sigma^p$) is independent of the spacetime coordinates $t$, $\vec x$ and $r$. As for the worldvolume gauge field, we take the ansatz
\begin{align}\label{gaugeFieldAnsatz}
A=A_t(r)dt.
\end{align}
In addition, we set the worldvolume scalar fields equal to zero whose holographic implication for the boundary field theory is that the probe charge carries are massless.

Substituting \eqref{embedding} and \eqref{gaugeFieldAnsatz} into \eqref{FlavorDBIAction}, and factoring from the action the infinite volume of $\mathbf{R}^{d+1}$ by defining   ${\cal S}=\left(\int dt\,dx^1 \cdots dx^d\right) S$, we obtain
\begin{align}\label{actionDensity}
S=-\mathfrak{N}\,\!\int\!dr\,g_{xx}^{q/2}\,\sqrt{|g_{tt}|g_{rr} - (\partial_r {A}_t)^{2}}\,,
\end{align}
with the overall factor $\mathfrak{N}$ defined as
\begin{align}
\mathfrak{N}& =T_q \int d\sigma^{d+2} \cdots \, d\sigma^p \nonumber\\
&= T_q \hbox{Vol}(\Sigma).
\end{align}
Now, varying the action (density) $S$ with respect to $A_t$ yields
\begin{align}\label{APrimeEqOne}
g_{xx}^{q/2} \frac{\partial_r {A}_t}{\sqrt{|g_{tt}|g_{rr} - (\partial_r {A}_t)^2}} = c\,,
\end{align}
where $c$ is a constant of integration.
Solving this equation for $\partial_r A_t$, we find
\begin{align}\label{APrimeEqTwo}
\partial_r A_t=c\frac{r^{-(1+\zeta)}}{\sqrt{r^{-2q\xi} + c^2}}\,,
\end{align}
where, without loss of generality, we have chosen the positive branch for $\partial_r A_t$. Also, for the ease of notation, we defined two new parameters $\zeta$ and $\xi$ which are related to $z$ and $\theta$ according to
\begin{align}
\zeta&\equiv z-2\theta/d, \qquad\qquad \xi\equiv 1-\theta/d\,.
\end{align}
Note that since in our discussion we are only considering those theories for which $\theta \leq d$ and $z\geq1$, one then simply deduces
\begin{align}
\zeta\geq\xi\geq 0.
\end{align}

Denoting the chemical potential for the charge density by $\mu$, it is given by the leading asymptotic behavior of $A_t$ near the boundary, namely
\begin{align}\label{chemicalPotentialDef}
\mu \equiv A_t (\epsilon \to 0),
\end{align}
where $\epsilon$ is a cutoff. Expanding \eqref{APrimeEqTwo} near $\epsilon\to 0$, one obtains
\begin{align}\label{asymptoticA}
A_t (\epsilon\to 0)=
\begin{cases}
\mu+\frac{c}{q\xi-\zeta} {\epsilon}^{q\xi-\zeta}+\cdots & q\xi\neq\zeta \, \\
\mu+ c\log \epsilon+\cdots & q\xi=\zeta
\end{cases} \,,
\end{align}
where the dots represent subleasing terms which would vanish in the $\epsilon\to 0$ limit.
The expansions in \eqref{asymptoticA} then imply that, up to some multiplicative factor, the charge density in the boundary field theory $\rho \equiv\langle J^t\rangle$ is simply given by $c$ provided that $q\xi > \zeta$. Notice that the case $\theta=d-1$ belongs to this category. (For $q\xi \leq \zeta$  the first term becomes subleading compared to the second term which presumably indicates that $c$ should be identified as the chemical potential). For technical reasons, we will exclude the case $q\xi=\zeta$ in our subsequent discussions.

To fully determine $\rho$  in terms of $\mu$, one should specify a boundary condition for $A_t$  in the IR. Imposing the IR boundary condition
\begin{align}
A_t(r\to \infty) = 0,
\end{align}
the boundary theory chemical potential can then easily be calculated by integrating the equation \eqref{APrimeEqTwo}, that is $\mu =\int^{\infty}_0 dr\, \partial_r A_t$, yielding
\begin{align}\label{muDef}
\mu= \frac{1}{2\sqrt{\pi}q\xi}\Gamma \left(\frac{\zeta}{2 q \xi}\right) \Gamma \left(\frac{1}{2}-\frac{\zeta}{2 q \xi}\right) c^{\frac{\zeta}{q \xi}}.
\end{align}

To determine other observables in the probe sector, such as the energy density, pressure, etc,  it is convenient to choose an ensemble. For example, in the grand canonical ensemble at zero temperature where one holds the chemical potential fixed, the energy density $\varepsilon$ and the pressure $p$ can then easily be computed from the Gibbs free energy density) $\Omega$, using the (zero-temperature) relations $\varepsilon =  \Omega +\mu \rho$ and $p=-\Omega$. In the context of the AdS/CFT correspondence, $\Omega$ is given by the negative of the renormalized on-shell action.
The on-shell action of the probe sector, which is obtained by inserting the solution for $\partial_r A_t$, given in the equation \eqref{APrimeEqTwo}, back into the action \eqref{actionDensity}, yielding
\begin{align}\label{onshellAction}
S_{\rm on-shell} =- \mathfrak{N}\!\int^{\infty}_0 dr\,\frac{r^{-(1+\zeta+2q\xi)}}{\sqrt{r^{-2q\xi} + c^2}},
\end{align}
is naively divergent in the boundary where $r\to 0$. Indeed, not surprisingly, the structure of the UV divergent terms in \eqref{onshellAction} is very similar to the divergent terms which appear in the studies of probe D-branes embedded in asymptotically Lifshitz spacetimes, as discussed in \cite{Hartnoll:2009ns}. Removing the divergences via analytic continuation and therefore, one can then easily verify that
\begin{align}\label{zeroTempEqs}
\rho=\mathfrak{N}\,c, \quad\,\, \varepsilon = \mathfrak{N}\frac{q\xi\,c}{q \xi+\zeta}\,\mu, \quad\,\, p= \mathfrak{N}\frac{\zeta c}{q \xi+\zeta}\,\mu,
\end{align}
with $c$ is given in terms of $\mu$ according to \eqref{muDef}. Also, using the above expressions, the the speed of first sound takes the form $v_1^2=(\partial p/\partial \varepsilon)_\mu=\zeta/q\xi$.

If a metric of the form \eqref{bckgr2} is viewed in its own right, and not as the IR limit of, say,  an asymptotically AdS solution, then the UV divergences could, in principle, be properly taken care of by first considering the system at a cutoff $r_\Lambda=\epsilon$, followed by adding appropriate local counterterms  and only then taking the $\epsilon \to 0$ limit, in a manner similar to the holographic renormalization prescription in asymptotically AdS spaces \cite{de Haro:2000xn,Bianchi:2001kw,Skenderis:2002wp, Karch:2005ms}  (although the structure of the divergent terms are quite different in these contexts\footnote{See also \cite{Gouteraux:2011qh} for a recent attempt in the holographic renormalization of Einstein-Maxwell-dilaton theories.}).
Although in section \ref{sec:3} we write down the appropriate counterterms needed in our calculations, we leave the issue of holographic renormalization in spacetimes of the form \eqref{bckgr2} for future studies.

\section{Current Two-Point Functions\label{sec:3}}

In this section, we study fluctuations of the worldvolume gauge field at linearized level. We then use the results and analytically calculate the low frequency and low momentum, behavior of the retarded correlators of the current operators in the boundary field theory. In this paper, however, we mainly focus on calculating these correlators in the longitudinal channel.  An analysis for the transverse currents can similarly be performed, and is left for future work.

\subsection{Linearized Equations of Motion}

In the following, we present  the linearized equations of motion for fluctuations of the worldvolume gauge field. We then solve them in the next section given some appropriate  boundary conditions. So, let us define
\begin{align}\label{fluctuation}
A_\mu= \bar A_\mu+ a_\mu,
\end{align}
where $\bar A_\mu = A_t(r)$ is the background profile of the gauge field, given in the equation \eqref{APrimeEqTwo}, and $a_\mu$ is the fluctuation. For later convenience, we also define
\begin{align}
f_{\mu\nu}= \partial_\mu a_\nu - \partial_\nu a_\mu.
\end{align}
We choose to work in the so-called radial gauge where we set $a_r = 0$.
To proceed, we find it convenient to Fourier transform the fluctuations to momentum space. Without loss of generality, we take the momentum vector to be in the $x^1$- direction. Denoting hereafter $k\equiv k^1$, we write
\begin{align}\label{FourierTrans}
a_\mu&\sim e^{-i\omega t} e^{ik x} a_{\mu}(r).
\end{align}

Substituting \eqref{FourierTrans} into the action \eqref{FlavorDBIAction} and keeping terms up to the quadratic order in fluctuations, one observes that the longitudinal fluctuations, $a_t$ and $a_x$, get decoupled from the rest of the fluctuations (\ie those in the transverse channel).  In coordinate space, the quadratic action for $a_t$ and $a_x$ takes the form
\begin{align}\label{quadraticAtAxAction}
\hskip-0.05inS_{(2)}&=\frac{\mathfrak{N}}{2}\!\int\!dr\,dt\,dx\,g_{xx}^{\frac{q}{2}} \left\{ \frac{g_{rr} {f}_{tx}^2 - |g_{tt}|(\partial_r{a}_x)^{2}}{g_{xx}\sqrt{|g_{tt}| g_{rr} - (\partial_r\bar{A}_t)^{2}}} \right.\nonumber\\
& \qquad\qquad\qquad\,\,\,\,\,\,\,\,\,\,\,\,+\left.\frac{|g_{tt}| g_{rr} (\partial_r{a}_t)^{2}}{\left[|g_{tt}| g_{rr} - (\partial_r\bar{A}_t)^{2} \right]^{3/2}} \right\},
\end{align}
where  $S_{(2)} = {\cal S}_{(2)}/ {\rm Vol}(\mathbf{R}^{p-2})$.
The equations of motion for $a_t$ and $a_x$  coming from the action \eqref{quadraticAtAxAction} then take the following form (in momentum space)
\begin{align}\label{atEq}
0&=\partial_r \left\{ \frac{g_{xx}^{q/2} |g_{tt}| g_{rr} (\partial_r a_t)}{\left[|g_{tt}| g_{rr} - (\partial_r\bar{A}_t)^{2} \right]^{3/2}} \right\} \nonumber\\
&\,\,\,\,\,\,\,- \frac{g_{xx}^{q/2-1} g_{rr}}{\sqrt{|g_{tt}| g_{rr} - (\partial_r\bar{A}_t)^{2}}}
k \left(k a_t + \omega a_x \right)  \,, \\ \nonumber\\
0&=\partial_r \left\{ \frac{g_{xx}^{q/2 - 1} |g_{tt}|(\partial_r a_x)}{\sqrt{|g_{tt}| g_{rr} - (\partial_r\bar{A}_t)^{2} }} \right\} \nonumber\\
&\,\,\,\,\,\,\,+ \frac{g_{xx}^{q/2-1} g_{rr}}{\sqrt{|g_{tt}| g_{rr} - (\partial_r\bar{A}_t)^{2}}}
\omega\left(k a_t+\omega a_x\right)\label{axEq}\,.
\end{align}
Also, in the gauge $a_r = 0$, $a_t$ and $a_x$ satisfy a constraint equation
\begin{align}\label{constraintEq}
\omega\,g_{rr} g_{xx}(\partial_r a_t) + k \left[ |g_{tt}| g_{rr} - (\partial_r\bar{A}_t)^{2} \right] (\partial_r a_x) = 0 \,,
\end{align}
which comes from varying the DBI action with respect to $a_r$. Note that not all the above three equations are independent. In fact, one can easily show that equations \eqref{constraintEq} and \eqref{axEq} imply the equation \eqref{atEq}. This means that it suffices to solve just the constraint equation together with, say,  the equation \eqref{axEq}.  Furthermore, written in terms of the gauge-invariant combination
\be
\mathscr E(r)= \omega a_x(r) + k a_t(r) ,
\ee
the equations of motion for $a_t$ and $a_x$ reduce to a single second-order equation for ${\mathscr E}(r)$:
\begin{align}\label{eomE}
0&= \partial_r^2{\mathscr E}+ \left [ \partial_r \ln \left(  \frac{g_{xx}^{(q-3)/2} g_{rr}^{-1/2}|g_{tt}|}{u
\left(k^2 u^2 - \omega^2\right)} \right) \right] \partial_r \mathscr E \nonumber\\
&- \frac{g_{xx}}{|g_{tt}|}
\left(k^2 u^2 - \omega^2 \right) \mathscr E \,,
\end{align}
where $u$, which is a function of $r$, is defined as follows
\begin{align}
u(r)= \sqrt{\frac{|g_{tt}| g_{rr} - (\partial_r\bar{A}_t)^{2}}{g_{rr}g_{xx}}}=\frac{ r^{-(z-1)} }{ \sqrt{1 + c^2 r^{2q\xi} }} \,.
\end{align}

Since we are interested in calculating the retarded correlators, we solve the equation \eqref{eomE} with in-falling  boundary condition in the IR as $r \to \infty$ \cite{Son:2002sd}. Note that we are only interested in solving this equation \eqref{eomE} at low frequency and low momentum (as compared to the chemical potential).

\subsection{Solution of the Linearized Equations}
To solve the equation \eqref{eomE} in the regime of interest, we use a matching technique. In the $r \rightarrow \infty$ limit, the equation takes the form
\begin{align}\label{deepIREEq}
\partial_r^2{\mathscr E}- \frac{1}{r} \left [  1+\zeta-4\xi\right] \partial_r{\mathscr E}+ \omega^2 r^{2(1+\zeta-2\xi)} {\mathscr E}= 0 \,.
\end{align}
The general solution of this equation is given in terms of Bessel functions. Imposing in-falling boundary condition in the $r \rightarrow \infty$ limit, one can then easily show that
\begin{align}\label{sol1}
{\mathscr E} &= C\left(\frac{\omega r^z}{2z}\right)^{\frac{1}{2}-\frac{\xi}{z}}H^{(1)}_{\frac{1}{2}-\frac{\xi}{z}}\!
\left(\frac{\omega r^z}{z}\right)\,,
\end{align}
is a solution of the equation \eqref{deepIREEq}. In the solution \eqref{sol1},  $H^{(1)}$ is the Hankel function of the first kind.

We then solve the equation \eqref{eomE} in the regime of low frequency and low momentum with $\omega r^z\ll 1$ and $kr\ll 1$ (while keeping the ratio $\omega/k^z$ fixed). This can be achieved by dropping the potential term in (\ref{eomE}) (that is,  the term proportional to ${\mathscr E}$), yielding
\begin{align}\label{lowEnergyEEq}
0&=\partial_r^2{\mathscr E} \nonumber\\
&\,\,\,\,- \left[ \frac{2 (\zeta +1)+\xi  (q-6)}{r} - \frac{u'}{u} \frac{3k^2 u^2 - \omega^2}{k^2 u^2 -\omega^2} \right] \partial_r{\mathscr E}.
\end{align}
Keeping in mind that we are excluding the case $q\xi= \zeta$, the general solution of \eqref{lowEnergyEEq} takes the form
\begin{widetext}
\begin{align}\label{sol2}
{\mathscr E}&=C_1+C_2\frac{r^{q\xi-\zeta}}{\sqrt{1+c^2 r^{2 q \xi}}}\left\{\frac{k^2}{q \xi-\zeta}\,\,\,\!_2F_1\left[1,-\frac{\zeta}{2 q \xi},\frac{3}{2}-\frac{\zeta}{2 q \xi},-c^2 r^{2 q \xi}\right]\right.\nonumber\\
&\hskip1.55in\left.-\frac{\omega^2 r^{2(z-1)} \left(1+c^2 r^{2 q \xi}\right)}{2+\zeta +(q-4)\xi}\,\,\,\!_2F_1\left[1,1-\frac{1}{q}+\frac{z}{2 q \xi},\frac{3}{2}-\frac{1}{q}+\frac{z}{2 q \xi},-c^2 r^{2 q \xi}\right]\right\},
\end{align}
\end{widetext}
where $_2F_1$ is the hypergeometric function.

The solutions \eqref{sol1} and \eqref{sol2} should then be matched in the region(s) where they overlap. For $\omega r^z\ll1$ and $z\neq2\xi$, the solution in (\ref{sol1}) reduces to
\begin{align}\label{exp1}
{\mathscr E}&\approx C\Bigg\{-\frac{i}{\pi}\Gamma\Big(\frac{1}{2} - \frac{\xi}{2+\zeta -2\xi}\Big)+\Bigg[\frac{1+i\tan\Big(\frac{\pi\xi}{2+\zeta -2\xi}\Big)}{\Gamma\left(\frac{3}{2}-\frac{\xi}{2+\zeta -2\xi}\right)}\nonumber\\
&\qquad \,\times \Big(\frac{\omega }{4+2\zeta-4\xi}\Big)^{1-\frac{2\xi}{2+\zeta -2\xi}}r^{\zeta -4 \xi +2}\Bigg]\Bigg\},
\end{align}
while for $z=2\xi$, it becomes
\be\label{exp2}
{\mathscr E}\approx C \left\{1+\frac{2 i}{\pi} \left[\log (\omega r^{2\xi})-\log \left(4 \xi\right)+\gamma \right]\right\},
\ee
with $\gamma$ being the Euler-Mascheroni constant. On the other hand, in the limit $r\to\infty$ and for $z\neq2\xi$, the solution (\ref{sol2}) reduces to
\begin{align}\label{exp1b}
{\mathscr E}&\approx\mathfrak{}C_1-C_2\frac{\omega^2}{c(2+\zeta-4\xi)}r^{2+\zeta-4\xi}\nonumber\\
&+\frac{C_2}{\sqrt{\pi}}\left[\frac{k^2c^{-1+\frac{\zeta}{q\xi}}}{q\xi}{\Gamma\Big(1+\frac{\zeta}{2q\xi}\Big)\Gamma\Big(\tfrac{1}{2}-\frac{\zeta}{2q\xi}\Big)}{}\right.\\
&\left.\qquad \,\,+\frac{\omega^2c^{-1+\frac{2}{q}-\frac{z}{q\xi}}}{2+\zeta-4\xi}{\Gamma\Big(1+\frac{1}{q}-\frac{z}{2q\xi}\Big)\Gamma\Big(\frac{1}{2}-\frac{1}{q}+\frac{z}{2q\xi}\Big)}\right],\nonumber
\end{align}
which nicely matches onto \eqref{exp1}. For $z=2\xi$, we obtain
\begin{align}\label{exp2b}
{\mathscr E}&\approx C_1-\frac{C_2\omega^2 }{c q\xi}\log\left(\frac{2c}{\omega^{q/2}}\right)-\frac{C_2\omega^2}{2c\xi}\log(\omega r^{2\xi})\\
&+\frac{C_2k^2}{\sqrt{\pi}q\xi}c^{-1+\frac{2}{q}-\frac{2\theta/d}{q\xi}}\Gamma\Big(\frac{1}{2}-\frac{2}{q}+\frac{1}{q\xi}\Big)\Gamma\Big(1+\frac{2}{q}-\frac{1}{q\xi}\Big).\nonumber
\end{align}
which matches onto  \eqref{exp2}.

Finally, to obtain the correlators we need the behavior of $\mathscr E$ in the $r\to 0$ limit. Expanding  the solution \eqref{sol2}
 in the $r\to 0$ region, we obtain
\begin{align}\label{asymptoticE}
\mathscr E\approx C_1+C_2\frac{k^2r^{q\xi-\zeta}}{q\xi-\zeta}.
\end{align}

\subsection{Correlation Functions\label{corrlfnc}}

In terms of $\mathscr E$, the quadratic action \eqref{quadraticAtAxAction} for $a_t$ and $a_x$ takes the form
\begin{align}\label{actionE}
S_{(2)} &= \frac{\mathfrak{N}}{2}\!\int\!dr \, d\omega \, dk \,\frac{1}{u}\sqrt{g_{xx}^{(q-3)} g_{rr}}\nonumber\\
&\qquad \qquad\quad\,\, \times \left [ \mathscr E^2 +\frac{|g_{tt}|}{g_{rr}(u^2 k^2 - \omega^2)} (\partial_r{\mathscr E})^2 \right] .
\end{align}
Introducing a cutoff at $r=\epsilon$ and integrating by parts, we obtain, in the $\epsilon\to 0$ limit,
\begin{align}\label{actionE}
S_{(2)} = -\frac{\mathfrak{N}}{2}\!\int\!d\omega \, dk \frac{\epsilon^{1+\zeta-q\xi}}{k^2} \mathscr E(\epsilon) \, \partial_r \mathscr{E}(\epsilon) \,.
\end{align}

To obtain the retarded two-point functions, one needs to functionally differentiate the properly renormalized on-shell action with respect to the sources. In our case, for the matrix of the retarded correlators, we have
\begin{align}\label{Gtt}
G^{tt}_R(\omega,k) &= \frac{\delta^2S_{(2)}}{\delta a_t(\epsilon)^2}=k^2 \,\Pi(\omega, k),\\
G^{xx}_R(\omega,k) &= \frac{\delta^2S_{(2)}}{\delta a_x(\epsilon)^2}= \omega^2 \,\Pi(\omega, k),\\
G^{tx}_R(\omega,k) &= \frac{\delta^2S_{(2)}}{\delta a_t(\epsilon)\delta a_x(\epsilon)}= \omega\,k \,\Pi(\omega, k),\label{Gxx}
\end{align}
with
\begin{align}\label{defPi}
\Pi(\omega,k) \equiv \frac{\delta^2S_{(2)}}{\delta \mathscr E(\epsilon)^2} \,.
\end{align}
Thus, in order to compute the correlation functions $G^{tt}_R(\omega,k)$, $G^{xx}_R(\omega,k)$ and $G^{tx}_R(\omega,k)$, all we need to do is to compute a single function $\Pi(\omega,k)$.
To determine $\Pi(\omega,k)$, we first substitute the boundary behavior of the solution for $\mathscr E$ that we have already calculated, as well as its derivative $\partial_r {\mathscr E}(\epsilon)$, into the action \eqref{actionE} to obtain
\begin{align}\label{onshellact}
S_{(2)} = -\frac{\mathfrak{N}}{2}\!\int\!d\omega  dk\,C_1C_2\,.
\end{align}
Note that for $q\xi< \zeta$, the action should be supplemented with appropriate boundary terms to cancel the divergences as $\epsilon\to0$.
To make $S_{(2)}$ finite, a boundary term of the form\footnote{Note that new (logarithmic) divergences also appear for the case $q\xi=\zeta$. Had we considered this particular case, the appropriate boundary terms for canceling the divergences would have taken the form
\begin{align}
S_{\epsilon}&=\frac{\mathfrak{N}}{2}\int d\omega d\k \frac{1}{k^2\log\epsilon}\sqrt{-\gamma}\,\gamma^{tt}\gamma^{xx} \mathscr E^2(\epsilon)\nonumber\\
&=\frac{\mathfrak{N}}{2}\int d\omega d\k \frac{1}{k^2\log\epsilon}  \mathscr E^2(\epsilon).
\end{align}}
\begin{align}
S_{\epsilon} &= \frac{\mathfrak{N}}{2} \int d\omega d\k \frac{q\xi-\zeta}{k^2}\sqrt{-\gamma}\, \gamma^{tt}\gamma^{xx} \mathscr E^2(\epsilon) \nonumber\\
&= \frac{\mathfrak{N}}{2} \int d\omega dk \frac{q\xi-\zeta}{k^2} \epsilon^{\zeta-q\xi} \mathscr E^2(\epsilon),
\end{align}
should be added for the case $q\xi<\zeta$.

In order to evaluate \eqref{defPi}, one needs to determine whether to differentiate the action with respect to $C_1$ or $C_2$. For $q\xi > \zeta$ the action should be differentiated with respect to $C_1$ in which case $C_2$ in then determined in terms of $C_1$ while for the case $q\xi < \zeta$ it is the other way around.  Suppose we take the derivate of $S_{(2)}$ with respect to $C_1$. Therefore, $C_2$ should be determined in terms of $C_1$. This can easily be done by matching the constant and the coefficient of the $r^{2+\zeta-4\xi}$ term (or logarithm) in the equations (\ref{exp1}) and (\ref{exp2}) with those in (\ref{exp1b}) and (\ref{exp2b}). Having done so, we differentiate the action \eqref{onshellact} twice with respect to $C_1$, ignore the overall factor $\mathfrak{N}$, and obtain the following structure for $\Pi(\omega, k)$,
\be\label{slnPi}
\Pi(\omega,k)\propto \frac{1}{k^2 - \alpha_1 \, \omega^2- \alpha_2 \,  \omega^2\,G_0(\omega)} \,,
\ee
where
\be\label{G0}
G_0(\omega) =
\begin{cases}
\omega^{-1+\frac{2\xi}{z}} & z\neq2\xi \, \\
\log\left({\alpha\omega^2}\right) & z=2\xi
\end{cases} \,,
\ee
with $\alpha_1$, $\alpha_2$ being some constants. Substituting \eqref{slnPi} into  \eqref{Gtt}--\eqref{Gxx}, one arrives at
\begin{align}\label{GttExpression}
G^{tt}_R(\omega,k) &\propto \frac{k^2}{k^2 - \alpha_1 \, \omega^2- \alpha_2 \,  \omega^2\,G_0(\omega)}\,,\\
G^{xx}_R(\omega,k) &\propto \frac{\omega^2}{k^2 - \alpha_1 \, \omega^2- \alpha_2 \,  \omega^2\,G_0(\omega)}\,,\\
G^{tx}_R(\omega,k) &\propto \frac{\omega k}{k^2 - \alpha_1 \, \omega^2- \alpha_2 \,  \omega^2\,G_0(\omega)}\,. \label{GtxExpression}
\end{align}
As for the constants $\alpha_1$, $\alpha_2$ and $\alpha$ appearing in \eqref{slnPi} and \eqref{G0}, one finds the expressions
\begin{align}\label{alphaOneZeroSound}
\hskip-0.2in\alpha_1&=\frac{c^{-\frac{2(z-1)}{q\xi}}\Gamma\left(\frac{1}{q}-\frac{z}{2q\xi}\right)\Gamma\left(\frac{1}{2}-\frac{1}{q}+\frac{z}{2q\xi}\right)}{2\,\Gamma\left(1+\frac{\zeta}{2q\xi}\right)\Gamma\left(\frac{1}{2}-\frac{\zeta}{2q\xi}\right)}\,,\\ \nonumber\\
\alpha_2&=\frac{i\pi^{-1/2}q\xi(2z)^{-\frac{2\xi}{z}}c^{-\frac{\zeta}{q\xi}}\Gamma\Big(\frac{1}{2}-\frac{\xi}{z}\Big)^2}
{\left[1+i\tan(\frac{\pi\xi}{z})\right]\Gamma\left(1+\frac{\zeta}{2q\xi}\right)\Gamma\Big(\frac{1}{2}-\frac{\zeta}{2q\xi}\Big)},\label{alphaTwoZeroSound}
\end{align}
for $z\neq2\xi$, whereas for $z=2\xi$ one obtains
\begin{align}\label{alphaOneLogCase}
\alpha_1&=\frac{q c^{-\frac{2(1-2\theta/d)}{q\xi}}\Gamma\left(\frac{1}{2}\right)}{2\,\Gamma\left(\frac{1}{2}-\frac{2}{q}+\frac{1}{q\xi}\right)\Gamma\left(1+\frac{2}{q}-\frac{1}{q\xi}\right)}\nonumber\\
&\times\left(\frac{i \pi}{2}+\log(4\xi)-\gamma\right)\,,\\
\alpha_2&=-\frac{q c^{-\frac{2(1-2\theta/d)}{q\xi}}\Gamma\left(\frac{1}{2}\right)}{4\,\Gamma\left(\frac{1}{2}-\frac{2}{q}+\frac{1}{q\xi}\right)\Gamma\left(1+\frac{2}{q}-\frac{1}{q\xi}\right)}\,,\\
\alpha &= (2c)^{-4/q}.\label{alphaLogCase}
\end{align}

\subsection{Poles of the Current Correlators }

From the denominator of the expressions \eqref{GttExpression}--\eqref{GtxExpression}, one observes that for $1\leq z<2\xi$ the term $\alpha_1\omega^2$ dominates over the term $\alpha_2\,\omega^2\,G_0(\omega)$ as $\omega\to 0$\footnote{The lower limit on $z$ comes from our earlier considerations. See the paragraph below \eqref{NEC}.}. This behavior results in a mode with a linear dispersion relation in the low energy spectrum.
To be more specific, let us investigate the behavior of the dominant pole and extract from it the advocated linear dispersion relation. First, note that the correlators  \eqref{GttExpression}--\eqref{GtxExpression} will have a pole whenever the following relation is satisfied:
\be\label{komega}
k(\omega)=\pm\sqrt{\alpha_1\,\omega^2 +\alpha_2\, \omega^2 G_0(\omega)}\,.
\ee
Next, to find the dispersion relation, $\omega =\omega(k)$,  the this expression needs to be inverted. In the following, we separately consider three different cases: $1\leq z<2\xi$,  $z>2\xi$, and $z=2\xi$

\subsubsection{$1\leq z<2\xi$: A zero sound-like mode}

As mentioned above, for $1\leq z<2\xi$, the term $\omega^2$ in the denominator of the expressions \eqref{GttExpression}--\eqref{GtxExpression} dominates over the term $\omega^2\,G_0(\omega)$ in the $\omega\to 0$ limit.  As a result the expression in \eqref{komega} could be expanded as follows
\begin{align}
k(\omega)=\pm\omega\sqrt{\alpha_1}\left[1+\frac{\alpha_2}{2\alpha_1}\omega^{-1+\frac{2\xi}{z}}+\mathcal{O}\Big(\omega^{-2+\frac{4\xi}{z}}\Big)\right].
\end{align}
This expression can then be inverted to give an expression for $\omega(k)$ as follows:
\begin{align}\label{zeroSoundPole}
\hskip-0.2in\omega(k)&=\pm\frac{k}{\sqrt{\alpha_1}}\left[1-\frac{\alpha_2}{2\alpha_1}\left(\frac{k}{\sqrt{\alpha_1}}\right)^{\frac{-1+2\xi}{z}}\right.\nonumber\\
&\left.\qquad\qquad\,\,\,\,\,\,\,\,+\mathcal{O}\left(\frac{k}{\sqrt{\alpha_1}}\right)^{\frac{-2+4\xi}{z}}\right].
\end{align}
For $1\leq z<2\xi$, $\alpha_1$ is real while $\alpha_2$ is complex.
Since ${\rm Im}\, \omega(k)\sim k^{2\xi/z}$ which is smaller than the real part ${\rm Re}\, \omega(k)\sim k$ as $k\to 0$, one deduces that this pole should represent a stable excitation. Moreover, we can easily see that, in the regime of parameters we are considering here the imaginary part is always negative. Such a linearly dispersing mode which appears as pole in the retarded correlators of the charge density operator in our zero-temperature finite density theories reminds one of the zero sound mode in Fermi liquids. In fact, the authors of \cite{Karch:2008fa} were first to realize the existence of such a zero sound-like mode in a holographic zero-temperature finite density field theory dual to a D3/D7 system. What makes their observation interesting is that this mode was found in a  boundary field theory which actually exhibits non-Fermi liquid behavior. (See also \cite{Kulaxizi:2008kv, Davison:2011ek} for related studies.) The existence of the holographic zero-sound mode was also shown in a number of other concrete holographic systems \cite{Kulaxizi:2008jx, HoyosBadajoz:2010kd, Lee:2010ez, Edalati:2010pn, Ammon:2012je}. In fact, its existence seems to go beyond the probe approximation of the charge sector and is even present \cite{Edalati:2010pn} in finite-density field theories dual to the (backreacted) extremal Reissner-Nordstr\"om-AdS backgrounds. Our finding in this section shows the existence of a similar mode in yet another finite-density holographic system.  Once combined with other studies in the literature, our observation could be viewed as providing further support for the genericness of the holographic zero sound mode in finite density holographic theories in the regime of zero or low temperature (compared to the chemical potential).

From \eqref{zeroSoundPole}, and using \eqref{alphaOneZeroSound}, the speed of this zero sound-like mode in our setup is given by
\begin{align}\label{zeroSoundVelocity}
v_0^2=2c^{\frac{2(z-1)}{q\xi}}\frac{\Gamma\left(1+\frac{\zeta}{2q\xi}\right)\Gamma\left(\frac{q\xi-\zeta}{2q\xi}\right)}{\,\,\,\Gamma\left(\frac{2\xi-z}{2q\xi}\right)\Gamma\left(\frac{1}{2}+\frac{z-2\xi}{2q\xi}\right)}\,.
\end{align}
Note that setting $z=1$, $\theta =0$ and $q=d$, the above expression reduces to the speed of the holographic zero sound found in \cite{Karch:2008fa}, while for $\theta=0$ (and finite $z\neq 1$) it matches the one in \cite{HoyosBadajoz:2010kd}.

\subsubsection{$z>2\xi$}

For this case, as $\omega \to 0$, the term $\omega^2G_0(\omega)$ in the denominator of our expressions for the correlators will dominate over the term $\omega^2$.  Expanding \eqref{komega} in this case, we then find
\begin{align}
k(\omega)&=\pm\omega^{\frac{1}{2}+\frac{\xi}{z}}\sqrt{\alpha_2}\left[1+\frac{\alpha_1}{2\alpha_2}\omega^{1-\frac{2\xi}{z}}\right.\nonumber\\
&\left.\qquad\qquad\qquad\,\,\,\,\,\,\,\,\,\, +\,\mathcal{O}\left(\omega^{2-\frac{4\xi}{z}}\right)\right],
\end{align}
which, in turn, results in
\begin{align}
\omega(k)&=\pm\left(\frac{k}{\sqrt{\alpha_2}}\right)^{\frac{2z}{z+2\xi}}\left[1-\frac{z}{z+2\xi}\frac{\alpha_1}{\alpha_2}\left(\frac{k}{\sqrt{\alpha_2}}\right)^{\frac{2z-4\xi}{z+2\xi}}\right.\nonumber\\
&\left.\qquad\qquad\qquad\qquad\,\,\,\,\,+\,\mathcal{O}\left(\frac{k}{\sqrt{\alpha_1}}\right)^{\frac{4z-8\xi}{z+2\xi}}\right].
\end{align}
Note that the leading term in the above expression for $\omega(k)$ is now complex.
This mode is gapless with a dispersion relation which is not linear (recall that we are considering the case $z>2\xi$). Moreover, although its imaginary part is negative, due to the fact that it is of the same order as the real part, this pole cannot be interpreted as representing a stable excitation.

\subsubsection{$z=2\xi$}

Finally, expanding \eqref{komega} for the case $z= 2\xi$, we obtain
\begin{align}
k(\omega)&=\pm\omega\sqrt{\alpha_2 \log(\alpha\, \omega^2)}\nonumber\\
&\,\,\,\,\,\times \left[1+\frac{\alpha_1}{2\alpha_2\log(\alpha\, \omega^2)}+\mathcal{O}\Big(\log(\alpha\, \omega^2)\Big)^{-2}\right].
\end{align}
From this expression, one again observes that $\omega \to 0$ as $k\to 0$, implying that the mode is gapless, although not linearly dispersing.

Note that in all of the above three cases the pole in the current correlators turned out to be  gapless (assuming $z$ is finite). In other words, at low energy as $\omega \to 0$, the mode scales toward the origin of the momentum. This kind of behavior is perhaps not surprising as it simply follows from the scaling transformation \eqref{scalings}, as pointed out in \cite{Hartnoll:2012wm}.

\subsection{Some Comments}

In this section we would like to make few comments regarding the low energy and low momentum behavior of the longitudinal current correlators and, in particular, the mode we found in the low energy spectrum.

First, in the class of holographic theories we are considering here, our calculations in the previous section shows that the retarded correlators of longitudinal currents exhibit markedly different behavior at low frequency and low momentum in the parameter space of the allowed values of $z$ and $\theta$. More specifically, we found that in the probe sector, theories for which $1\leq z<2(1-\theta/d)$ support a stable linearly-dispersing excitation (a zero sound-like mode) in the low energy spectrum  whereas for $z\geq 2(1-\theta/d)$ the pole does not correspond to a stable quasiparticle excitation. As far as the change in the low energy behavior of two-point functions is concerned, the situation here is similar to \cite{Edalati:2012tc} where the response of these holographic theories to a disturbance caused by a massive charged particle (represented in the bulk by a fundamental string) was studied. It was shown in \cite{Edalati:2012tc} that  the two-point functions of the quantum fluctuations of the massive probe exhibit different characteristics at low energies depending on whether the dynamical critical exponent $z$ is greater, or less than $2(1-\theta/d)$. More specifically, for $z > 2(1-\theta/d)$, the aforementioned two-point functions have no memory of the inertial mass of the massive charged particle at low energies. Also, as shown in \cite{Dong:2012se}, the two-point functions of scalar operators dual to massive scalar fields in the bulk exhibit a transition from a universal power law at short distances to a nontrivial exponential behavior at long distances in theories with $\theta>0$.

Second, the observation that for $z\geq 2(1-\theta/d)$ the charge density correlators does not exhibit a sharp zero sound-like peak poses the following question.   What happens to the fate of this mode as one crosses the line of $z=2(1-\theta/d)$ in the parameter space of $z$ and $\theta$? One possibility is that for $z\geq 2(1-\theta/d)$ there might be decoherent states at low energies which interact rather strongly with this mode,  resulting in the destabilization of an otherwise sharp linearly-dispersing excitation. Investigating this possibility requires the analysis of the low energy behavior of other correlators in the theory, such as the correlators of transverse currents. This mechanism would be similar to what was observed in \cite{Edalati:2010ww, Edalati:2010ge} for a different class of finite density holographic theories \cite{Lee:2008xf, Liu:2009dm, Cubrovic:2009ye, Faulkner:2009wj} where by changing some parameters, the well-defined excitations of a Fermi surface interact with a continuum of decoherent low energy states and become destabilized.

Finally, as we mentioned in the introduction, bulk gravitational theories whose metric is of the form \eqref{bckgr2} exhibit an intriguing feature. For $\theta=d-1$, the holographic computation of the entanglement entropy, using the Ryu-Takayanagi formula \cite{Ryu:2006bv, Ryu:2006ef}, reveals a logarithmic violation\footnote{Note that in the holographic calculations of the entanglement entropy performed in \cite{Ogawa:2011bz, Huijse:2011ef}, in order to obtain a logarithmic violation of the area law it is only important that the metric of the background is of the form \eqref{bckgr2} with $\theta =d-1$, and the profile of the other bulk fields do not enter the computation. In particular, it is not important whether the charge degrees of freedom in the boundary theory comes from a U(1) bulk gauge field in the context of effective Einstein-Maxwell-theories, or by an embedded flavor D-brane.} of the area law in such a way that it is reminiscent of theories with Fermi surfaces  \cite{Ogawa:2011bz, Huijse:2011ef}. As a result, these bulk solutions have been proposed as potential gravity duals of field theories with Fermi surfaces even though there are no explicit fermions in the bulk. Thus, it is  interesting to explore other fermionic aspects, if any, of this particular class of theories. One such exploration is the fate of the holographic zero sound in this class of theories. After all,  at least in Fermi liquids, a zero sound is nothing but a collective excitation resulting in the oscillations of the Fermi surface.

As we argued above, there is a zero sound-like mode in the class of boundary theories where the dynamical critical exponent and the hyperscaling violation exponent satisfy the condition $1\leq z<2(1-\theta/d)$.  Now, since $d\geq 2$, the case $\theta = d-1$ would satisfy this condition only for $z=1$.  But (given that we are  assuming $\theta \leq d$), theories with  $\theta =d-1$ and $z=1$ which would have exhibited a holographic zero sound mode does not actually satisfy the null energy condition given in \eqref{NEC}. In fact, the null energy condition \eqref{NEC}, in conjuction with our further assumption $\theta \leq d$, implies that in theories with $z=1$ the hyperscaling violation exponent must satisfy $\theta \leq 0$. Thus, we conclude that sensible holographic theories with $\theta=d-1$ does not exhibit a holographic zero sound.\footnote{Recently, it was argued in \cite{Fan:2013mma} that theories with $\theta=d(d-1)/(d+1)$ also lead to a logarithmic violation of the area law in the entanglement entropy computation. For this set of parameters one can see that $d=2$ is the only value for which a sound mode can be found.} This conclusion may complicate the identification of this class of solutions (those with $\theta=d-1$) as gravity duals for field theories with Fermi surfaces. Our result on the fate of the holographic zero sound mode in finite density theories with $\theta=d-1$ could in principle be further investigated by computing the low energy spectral density of transverse current correlators at finite momentum  to see whether there the type of a $2k_{\rm F}$ singularity. In fact, using a bulk Einstein-Maxwell-Dilaton theory,  it has been explicitly shown in \cite{Hartnoll:2012wm} that the transverse current correlators in these holographic theories do not support any low-energy excitations at finite momentum for any finite $z$, a behavior which is different from what one typically expects from a system with fermionic nature.

\subsection{A Scaling Limit}
Before moving on to the next section, we would like to take a special  limit of our results to mainly compare them with the findings in \cite{Pang:2013ypa}. This limit involves  taking $z\to\infty$ while keeping the ratio $\eta=-\theta/z>0$ fixed. It is then easy to show that upon taking this limit,  the metric \eqref{bckgr2} becomes conformal to AdS$_2\times \mathbb{R}^{d}$. The dual theories in this case represent a class of semi-local quantum liquids \cite{Hartnoll:2012wm} with some appealing features.  For example, unlike some other holographic setups, the entropy density in such theories vanishes at low temperatures as $s \sim T^\eta$. Also, the spectral density of transverse currents is not exponentially suppressed at nonzero momentum, presumably indicating some fermionic nature of these theories.

Now, in this limit, the function $G_0(\omega)$ which appears in \eqref{slnPi} takes the form
\be
G_0(\omega) =
\begin{cases}
\omega^{-1+\frac{2\eta}{d}} & d/\eta\neq2 \, \\
\log\left({\alpha\omega^2}\right) & d/\eta=2
\end{cases} \,.
\ee
From this expression, it follows directly that there is a zero sound-like mode when $d/\eta<2$ but not otherwise, as pointed out in \cite{Pang:2013ypa}. Moreover, the expression \eqref{zeroSoundVelocity} for the zero-sound velocity obtained here coincides with the result of \cite{Pang:2013ypa} in this limit.

\section{An Effective Description\label{sec:4}}

Generalizing the argument of \cite{Nickel:2010pr}, in this section we write down an effective action for the zero sound-like that we found in the range $z <2(1-\theta/d)$ by interpreting it as a Goldstone boson arising from a specific symmetry breaking pattern. As advocated in \cite{Nickel:2010pr}, such a linearly-dispersing mode can be interpreted as a Goldstone boson, but not of the breaking of a global U(1) symmetry, but rather of the
breaking of a U(1)$_{\rm global}\times$U(1)$_{\rm gauge}$ symmetry
down to a diagonal U(1).  The imaginary part in the dispersion then arises as a result of the coupling of this dynamical U(1) field to a sector in the IR. In fact, for the case $z>2(1-\theta/d)$ where we did not find a zero sound-like mode, this IR sector would dominate the physics and effectively damps out an otherwise would-be a sharp  mode in the spectrum\footnote{We would like to thank Dam T. Son for a discussion on this point.}.

In the following, we are interested in the dynamics of the theory at distances larger than some scale. More specifically, we would like to obtain a low-energy effective Wilsonian description in the same spirit as in \cite{Heemskerk:2010hk,Faulkner:2010jy}. To do so, we choose a radial cutoff $r_\Lambda$ as representing a scale which roughly separates the UV part of the geometry (which is integrated out) from an IR region that encodes the dynamics at low energies.
The degrees of freedom of the theory can then be broken into the IR degrees of freedom, denoted collectively by $\phi_{\rm IR}$, which live in the $r>r_\Lambda$ part of the spacetime geometry, an  ``emergent gauge field'' denoted by $a_\mu=A_\mu(r= r_\Lambda)$ living on the radial slice $r=r_\Lambda$and the UV degrees of freedom $\phi_{\rm UV}$, which live in the $r<r_\Lambda$ region as well as the U(1) gauge field $A_\mu=A_\mu(r=0)$ at the boundary. The degrees of freedom in the IR will naturally couple to $a_\mu$ which serves as a source for the IR theory. The UV degrees of freedom couple to both $A_\mu$ and $a_\mu$. Thus, the partition function of the
theory in the presence of external fields, can be written as:
\begin{equation}
  \mathcal{Z}[A_\mu] = \int\!\mathcal{D}a_\mu\,\mathcal{D} \phi_{\rm UV}\, \mathcal{D}\phi_{\rm IR}\,
  e^ {iS[A_\mu,a_\mu,\phi_{\rm UV},\phi_{\rm IR}]},
\end{equation}
where
\begin{align}
 S[A_\mu,a_\mu,\phi_{\rm UV},\phi_{\rm IR}] &= S_{\rm UV}[A_\mu,\,a_\mu,\phi_{\rm UV}]\nonumber\\
&\,\,\,\,\,+ S_{\rm IR}[a_\mu,\phi_{\rm IR}].
\end{align}
The key point observed \cite{Nickel:2010pr} is that the UV theory is a confining theory
and can be rewritten as a theory of mesons.
At low energies, the only important one is precisely the Goldstone boson arising from the
breaking of U(1)$_{\rm global}\times$U(1)$_{\rm gauge}$ symmetry down to the diagonal U(1) group. If one fixes the values of the temporal and spatial
components of the gauge field $A_\mu$ ($\mu\neq r$) on the two
boundaries, then the Goldstone boson is given by Wilson line
\begin{align}
\phi = \int_{0}^{r_\Lambda} dr\, A_r(r,x).
\end{align}
However, in the radial gauge where one sets $A_r=0$, the boundary condition $A_\mu=0$ (for $\mu\ne r$) cannot be imposed at both boundaries.  If $A_\mu=0$ on one boundary, then it should be $A_\mu=\partial_\mu\phi$ on the other. The Goldstone boson in this case is just the gauge parameter $\phi$.

Given the symmetries, we can write down the action
\begin{align}\label{S-gauge-init}
S_{\rm UV} &= \frac{1}{2} \int\!d^{q+1}x\,\!\left[f_t^2 \left(\partial_t\phi-A_t+a_t\right)^2\right.\nonumber\\
 &\left.\qquad\qquad\qquad- f_s^2 \left(\partial_i\phi - A_i + a_i\right)^2\right],
\end{align}
where $f_t$ and $f_s$ are some low-energy constants to be determined. The holographic zero sound mode may then be interpreted within this framework as a mode coming from an excitation of the field $\phi$, and therefore its speed is given by the expression $v_0^2=f_s^2/f_t^2$. To find the value of $f_t^2$ and $f_s^2$, one should then match this effective field theory with holographic calculations.  If we freeze the Goldstone
boson to $\phi=0$ and turn on constant external fields $A_0$ and $A_i$, then
the coefficients $f_t^2$ and $f_s^2$ are obtained by expanding $S$ to
quadratic order in the external fields,
\begin{equation}
S = \frac{1}{2}( f_t^2A_t^2 - f_s^2 A_i^2) .
\end{equation}
Freezing the Goldstone boson at $\phi=0$ corresponds to working in the
radial gauge $A_r=0$ and putting $A_\mu=0$ at the horizon.

Let us consider all the fields to be only a function of the radial coordinate $r$, which means that the transverse fluctuations will decouple, and we can set them all equal to  zero. Then, up to quadratic order, the DBI action reduces to
\begin{align}
S&=\frac{\mathfrak{N}}{2}\!\int\!dr \, dt \, d^qx \,r^{z+1-q+(q-2)\theta/d}\nonumber\\
&\qquad \qquad\qquad \times \left[f^3(r)A'_t\,\!^2-r^{2(1-z)}f(r)A'_i\,\!^2\right]\,,
\end{align}
with $f(r)=(1+c^2r^{2q\xi})^{1/2}$. The equations of motion for $A_t$ and $A_i$ coming from the above action are then
\begin{align}
0&=\partial_r\left(r^{z+1-q+(q-2)\theta/d}f^3(r)A'_t\right)\,,\\
0&=\partial_r\left(r^{3-z-q+(q-2)\theta/d}f(r)A'_i\right)\,.
\end{align}
Solving these equations and substituting the solution back into the action, we then find
\begin{align}
f_t^2&=\mathfrak{N}\left[\int_0^{r_\Lambda}\!\frac{dr}{r^{z+1-q+(q-2)\theta/d}f^3(r)}\right]^{-1}\nonumber\\
&=\mathfrak{N}\frac{\sqrt{\pi } q \xi c^{1-\frac{2}{q}-\frac{(z-2)}{q \xi}}}{\Gamma \left(\frac{1}{2}-\frac{1}{q}-\frac{z-2}{2 q \xi}\right) \Gamma \left(1+\frac{1}{q}+\frac{z-2}{2 q \xi}\right)}\,,\\ \nonumber\\
f_s^2&=\mathfrak{N}\left[\int_0^{r_\Lambda}\!\frac{dr}{r^{3-z-q+(q-2)\theta/d}f(r)}\right]^{-1}\nonumber\\
&=\mathfrak{N}\frac{\sqrt{\pi} \left(z+(q-2)\xi\right) c^{1-\frac{2}{q}+\frac{z}{q \xi}}}{\Gamma \left(\frac{1}{q}-\frac{z}{2 q\xi}\right) \Gamma \left(\frac{3}{2}-\frac{1}{q}+\frac{z}{2 q \xi}\right)}\,,
\end{align}
where we have taken the limit $r_\Lambda\to\infty$. Finally, dividing the two expressions above  one easily obtains the velocity of the zero sound-like mode reported in the previous section in \eqref{zeroSoundVelocity}.

\section{Final remarks\label{sec:5}}

In this paper we studied longitudinal current correlators of fundamental matter probing strongly coupled quantum critical points with dynamical exponent $z$ and hyperscaling violation exponent $\theta$. We showed that the low energy behavior of these correlators exhibit rather totally different behavior as a function of $z$ and $\theta$. In particular, we found that for theories satisfying $z<2(1-\theta/d)$ there exists  a linearly-dispersing pole, to some extent reminiscent of the zero sound mode in Femi liquids. Furthermore, with some assumptions having to do with the sensibility of bulk gravitational theories, we argued that the set of parameters for which such a zero sound-like mode exists does not include the case $\theta=d-1$ where the entanglement entropy exhibits a logarithmic violation of the area law. Applying the argument of \cite{Nickel:2010pr} for our holographic setup, we also showed that such a mode could be given an effective description in terms of a Goldstone boson arising from the breaking of a U(1)$_{\rm global}\times$U(1)$_{\rm gauge}$ symmetry down to a diagonal U(1).

An issue we did not address in this paper was whether the zero sound-like mode that we found here for theories with $z<2(1-\theta/d)$ survives once backreaction is taken into consideration. In particular, the metric \eqref{bckgr2} would get modified in the $r\to 0$ region mainly due to the fact that the profile of a dilationic scalar blows up in that region. In the IR, as $r\to \infty$, the metric \eqref{bckgr2} would also get modified both from the flavor D-brane and the profile of other bulk fields, such as the diatonic scalar, although such IR modifications could be tamed by turning on a small temperature. As a first step towards addressing the issue of backreaction one could repeat the computation performed here for the case where the bulk solution represents an RG  flow from, say, an asymptotically AdS spacetime in the UV to a metric of the form \eqref{bckgr2} for a wide range of scales, and eventually to some other fixed point deep in the IR, as in \cite{Ogawa:2011bz, Bhattacharya:2012zu, Kundu:2012jn}.

Finally, a number of possibilities for the extension of our analysis can be considered. One possibility which would shed further light on the results of this paper is to analyze the behavior of current correlators in the transverse channel at low energy and both low and finite momentum.  Another  possible extension of our work is to consider the inclusion of a magnetic field along the lines of \cite{Goykhman:2012vy}.
Last, but not least, would be the analysis of the finite temperature behavior of the zero sound-like mode that we found, in a manner similar to \cite{Davison:2011ek}. Of particular interest would be the fate of this mode as it crosses the collisionless regime to a hydrodynamic regime.

\section*{Acknowledgements}

We would like to thank D. T. Son and W. Fischler  for helpful discussions.
This material is based upon work supported by the National Science Foundation under Grant No. PHY-0969020 and by the Texas Cosmology Center.

\end{document}